\documentclass[12pt]{article}
\usepackage{etex}
\usepackage{epsfig}
\usepackage{epsf}
\usepackage{latexsym}
\usepackage{amsfonts,amssymb,amsmath} 
\usepackage{mathtools}
\usepackage{color}
\usepackage[a4paper,vmargin=3cm,hmargin=2.1cm]{geometry}
\usepackage{tikz}
\epsfverbosetrue
\newcommand{\erf}{\text{\rm{erf\,}}}

\newcommand{\de}{\hbox{\rm{d}}}

\newcommand{\lb}{\left[}
\newcommand{\rb}{\right]}
\newcommand{\lp}{\left(}
\newcommand{\rp}{\right)}

\newcommand{\dpp}{\vcentcolon}
\newcommand{\bb}{\begin{eqnarray}}
\newcommand{\ee}{\end{eqnarray}}
\newcommand{\eee}{\nonumber\end{eqnarray}}
\newcommand{\qq}{\quad}

\begin{document}

\thispagestyle{empty}

\begin{center}
${}$
\vspace{3cm}

{\Large\textbf{On a sneezing universe}} \\

\vspace{2cm}

{\large
Andr\'e Tilquin\footnote{CPPM, Aix-Marseille University, CNRS/IN2P3, 13288 Marseille, France
\\\indent\qq supported by the OCEVU Labex (ANR-11-LABX-0060) funded by the
"Investissements d'Avenir" 
\\\indent\qq
French government program
\\\indent\qq
 tilquin@cppm.in2p3.fr },
Thomas Sch\"ucker\footnote{
CPT, Aix-Marseille University, Universit\'e de Toulon, CNRS UMR 7332, 13288 Marseille, France
\\\indent\qq supported by the OCEVU Labex (ANR-11-LABX-0060) funded by the
"Investissements d'Avenir" 
\\\indent\qq
French government program
\\\indent\qq
thomas.schucker@gmail.com }
}
 
\vspace{2cm}

\hfill{\em To the memory of Daniel Kastler and Raymond Stora}

\vspace{2cm}

{\large\textbf{Abstract}}

\end{center}

A 1-parameter class of quadratic equations of state is confronted with the Hubble diagram of supernovae. The fit is found to be as good as the one using the standard $\Lambda $CDM model. However this quadratic equation of state precludes objects with redshifts higher than $z_{\rm max}\sim 1.7$. Adding a fair amount of cold baryons to the model increases $z_{\rm max}$ without spoiling the fit.

\vspace{2.5cm}

\vskip 1truecm

\eject

\section{Introduction}

One attempt at easing the persisting tension between cosmic observations and theory is to admit an exotic matter component  with an ad hoc equation of state expressing the pressure $p$ of the component as a function of its energy density $\rho $:
\bb p= f(\rho )\label{baro}.\ee
A popular example is the Chaplygin gas \cite{chap} and generalisations with $f(\rho )= p_0+ w\,\rho +\alpha /\rho $. Another functional class, quadratic equations of state, $f(\rho )= p_0+ w\,\rho +\alpha \,\rho^2 $, has recently attracted attention: 

Using dynamical systems theory
Ananda \& Bruni \cite{ana} classify the many different behaviours of the Robertson-Walker and Bianchi I universes resulting from quadratic equations of state. Linder \& Scherrer \cite{lin} analyse asymptotic past and future evolution of universes with ``barotropic fluids'' i.e. matter with an equation of state (\ref{baro}).  

Motivated by Bose-Einstein condensates as dark matter, Chavanis \cite{chav} has written a very complete series of papers on equations of state with $f(\rho )= w\,\rho +\alpha \,\rho^q $ in Robertson-Walker universes including analytical solutions of the Friedman equations and connections with inflation.

Bamba et al. \cite{odin} classify possible singularities of Robertson-Walker universes in presence of exotic matter, in particular with quadratic equations of state. Reddy, Adhav \& Purandare \cite{redd} solve the Friedman equations analytically in Bianchi I universes with equations of state $f(\rho )= -\rho +\alpha \,\rho^2 $.
Singh \&  Bishi \cite{Singh} consider the same setting in some $f(R,T)$ modified gravity theories.

 In the recent reference \cite{sha} Sharov confronts Friedman universes with curvature and general quadratic equations of state to the supernovae and baryon acoustic oscillation data. 
 
 We would like to contribute to this discussion and concentrate on the particular case $p_0=0$ and $w=-1$. We justify our choice by Baloo's mini-max principle: a maximum of pleasure with a minimum of effort.

\section{Friedman's equations}

To start, let us write the Friedman equations without cosmological constant and without spatial curvature:
\begin{align}
3\,H^2&=8\pi G\,\rho ,\label{first}\\
2\,H'+3\,H^2&=-8\pi G\,p.\label{second}
\end{align}
The prime denotes derivatives with respect to cosmic time $t$. We set the speed of light to one and have the scale factor $a(t)$ carry dimensions of time. The Hubble parameter  is as usual $H\dpp=a'/a$. We can trade the second Friedman equation (\ref{second}) for the continuity equation,
\begin{align}
\rho '=-3H\,(\rho +p).\label{cont}\end{align} 
The coefficient $\alpha $ in the quadratic equation of state has dimensions and it will be convenient to write it in the form,
\begin{align}
p=-\rho +\alpha \,\rho ^2= -\rho +{\textstyle\frac{1}{4}}\,8\pi G\,\tau^2  \,\rho ^2,\label{state}\end{align}
where $\tau$ is a characteristic time.
Using this equation of state, the continuity equation (\ref{cont}) integrates readily:
\begin{align}
\rho =\,\frac{1}{8\pi \,G}\,\frac{3\,H_0^2}{1+{\textstyle\frac{9}{4}}\,H_0^2\tau^2\ln(a/a_0) }\, ,\end{align}
with integration constant $H_0\dpp=H(t_0)$, the Hubble parameter today which is related to the density today and $a_0\dpp=a(t_0)$ is the scale factor today, that without loss of generality can be set to $a_0=1$s in flat universes.
For vanishing $\tau$, we retrieve of course the cosmological constant $\Lambda =\rho /(8\pi G)$. 

The pleasure continues and the first Friedman equation (\ref{first}) integrates as easily:
\begin{align}
a=a_0\,\exp\,\frac{(t-t_s)^{2/3}-(t_0-t_s)^{2/3}}{\tau^{2/3}}\, ,\label{asneeze}
\end{align}
where $t_s$ is another integration constant. This scale factor agrees with the one obtained by Chavanis in appendix A of his second paper in reference \cite{chav}.

\begin{figure}[h]
\begin{center}
\includegraphics[width=14.5cm, height=6.5cm]{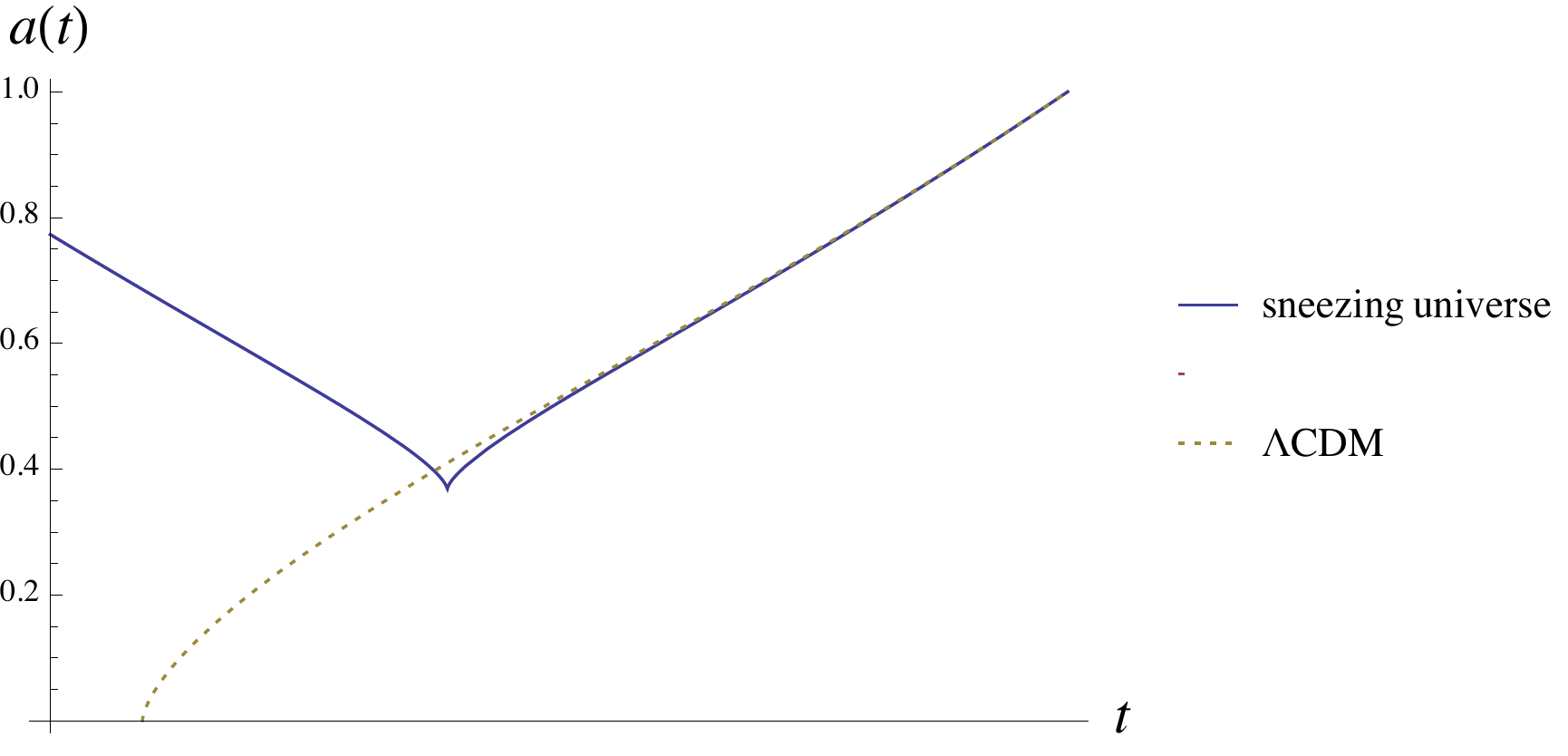}
\caption{Two scale factors  \label{2a}
 }
\end{center}
\end{figure}

Figure \ref{2a} shows this scale factor together with the one of the $\Lambda $CDM universe,
\begin{align}
a_{\Lambda {\rm CDM}}\,=\,\lp\frac{\cosh[\sqrt{3\Lambda }\,(t-t_{bb})]-1}{\cosh[\sqrt{3\Lambda }\,(t_0-t_{bb})]-1}\rp^{1/3}\, ,
\end{align}
with $t_{bb}$ the time when the big bang occurs.
 The fit is in anticipation of section \ref{versus}.    
 \medskip

 Already now, the scale factor (\ref{asneeze}) tells us a funny story: 
\begin{itemize}\item
no violent, hot bang, just a little sneeze at $t=t_s$. There, the scale factor remains positive and the temperature finite, only the Hubble parameter diverges together with the density and pressure of our exotic fluid. In the wording of reference \cite{odin} this singularity is of type III.
\item
After the sneeze, the universe expands with deceleration until $t_\pm=t_s\,+\,\tau/2^{3/2}$ where the deceleration parameter changes sign. Then the expansion accelerates for ever causing an event horizon.
\item
The sneeze is a mild singularity, i.e. integrable and -- if its finite temperature is sufficiently low -- transparent to light so that we can observe its past. There the universe contracts up to arbitrarily negative times, however with a particle horizon.
\item
The scale factor is invariant under time reversal with respect to the time of sneeze $t_s$.
\end{itemize}
We are sufficiently intrigued by this universe to try and confront it with supernova data.

\section{Hubble diagram}

For the sneezing universe, the Hubble constant is
\begin{align}
H_0=\,\frac{2}{3} \,\frac{(t_0-t_s)^{2/3}}{\tau^{5/3}}\, , 
\end{align}
the redshift is given by
\begin{align}
z+1=\,\frac{a_0}{a}\, =\,\exp-\,\frac{(t-t_s)^{2/3}-(t_0-t_s)^{2/3}}{\tau^{2/3}},\end{align}
and the apparent luminosity is
\begin{align}
\ell=\,\frac{L}{4\pi \,a_0^2\,\chi ^2}\,\frac{a^2}{a_0^2}\,,   
\end{align}
with the absolute luminosity $L$ and the dimensionless comoving geodesic distance
\begin{align}
\chi(t)\dpp =& \int_t^{t_0}\de\tilde t/a(\tilde t)\\[2mm]
=&\,{\textstyle\frac{3}{2}} \,\frac{\tau}{a_0}\, \exp R^2\lb-x^{1/3}\exp -x^{2/3}+{\textstyle\frac{1}{2}} \sqrt{\pi}\,\erf x^{1/3}\rb_{(t-t_s)/\tau} ^{(t_0-t_s)/\tau}\\[2mm]
=&\,{\textstyle\frac{3}{2}} \,\frac{\tau}{a_0} \lb
-R +{\textstyle\frac{1}{2}} \sqrt{\pi }\,\exp R^2 \,\erf R+
\exp R^2\,\lp \frac{t-t_s}{\tau}\rp^{1/3}\exp -\lp \frac{t-t_s}{\tau}\rp^{2/3} \right.\nonumber\\
&\qq\qq\qq\left.-{\textstyle\frac{1}{2}} \sqrt{\pi }\,\exp R^2 \,\erf \lp \frac{t-t_s}{\tau}\rp^{1/3}\rp,\label{only}
\end{align}
with the abbreviation
\begin{align} R\dpp=\lp \frac{t_0-t_s}{\tau}\rp^{1/3} \label{rdef}
\end{align} 
and the error function 
\begin{align}
\erf x\dpp=\,\frac{2}{\sqrt{\pi }}\int_0^x\exp -y^2\,\de y.
\end{align}
Note that the formula (\ref{only}) is a priori valid only for emission times $t\in [t_s,t_0]$. However due to the time reversal symmetry of the scale factor with respect to $t_s$, equation (\ref{only}) is valid for all $t$. In particular we have
\bb \chi (t)= 2\chi (t_s)-\chi (2t_s-t).\label{reversal}\ee
 \begin{figure}[h]
\begin{center}
\includegraphics[width=7.5cm, height=6.5cm]{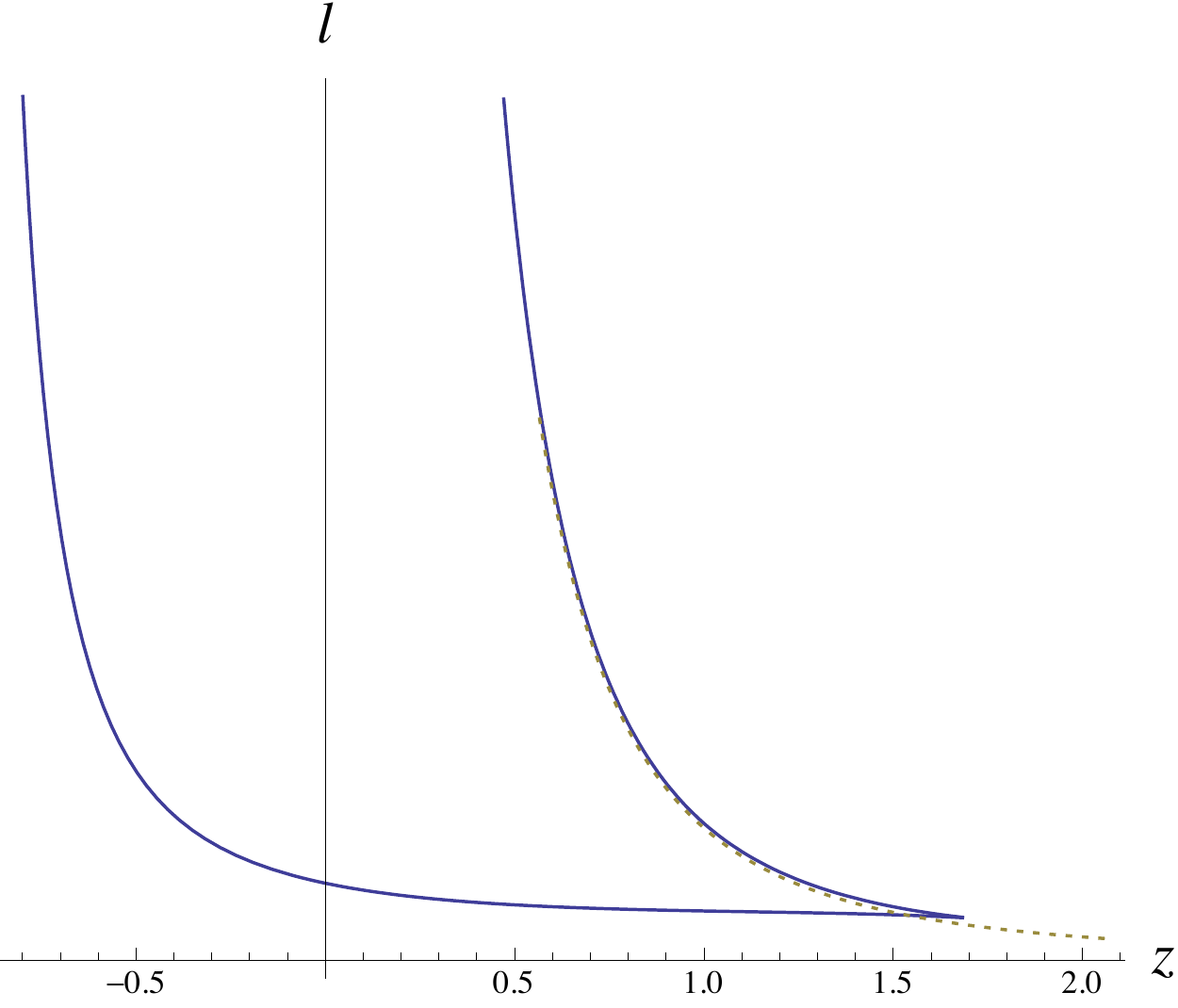}
\caption{Hubble diagrams of sneezing and $\Lambda $CDM universes, colour codes as in figure \ref{2a}. 
\label{2ell} }
\end{center}
\end{figure}
Note that the redshift is not an invertible function of time, it has a maximum:
\begin{align}
z_{\rm max}=z(t_s)=\,\exp\,\frac{(t_0-t_s)^{2/3}}{\tau^{2/3}}-1, \label{zmax}
\end{align}
and takes negative values, blueshift, for $t<2t_s-t_0$. Its minimal value $z_{\rm min}=-1$ occurs at $t=-\infty$ corresponding to  the particle horizon of finite comoving geodesic distance
\begin{align}
\chi _{\rm horizon}=\chi (-\infty)
=\,{\textstyle\frac{3}{2}} \,\frac{\tau}{a_0} \lb
-R +{\textstyle\frac{1}{2}} \sqrt{\pi }\,\exp R^2 \,\{\erf R+1\}
 \rb.
\end{align}
At this blueshift, $z_{\rm min}=-1$, the apparent luminosity tends to infinity because the energy of the arriving photon is boosted during the long contraction phase.

Figure \ref{2ell} shows the Hubble diagrams of the sneezing and the $\Lambda $CDM universes.

Because of the non-invertibility of the function $z(t)$ the Hubble diagram is a genuine parametric plot containing two functions $\ell_{1,2} (z)$,
\begin{align}
\ell_{1,2}(z)=\,\frac{L}{9\pi \, \tau^2}\,\frac{1}{(z+1)^2\,I_{1,2}^2(z)}\, \label{luminosity}.
\end{align}
The function $I_1(z)$ comes from emission times $t\in[t_s,\,t_0)$ corresponding to $z\in(0,\,z_{\rm max}]$, 
\begin{align}
I_1(z)=&\int_0^z\sqrt{R^2-\ln (\tilde z+1)}\,\de\tilde z\\
=&\,-R +{\textstyle\frac{1}{2}} \sqrt{\pi }\,\exp R^2\,\erf R\nonumber\\
&+(z+1)\, \sqrt{R^2-\ln ( z+1)}-{\textstyle\frac{1}{2}} \sqrt{\pi }\,\exp R^2\,\erf \sqrt{R^2-\ln ( z+1)}\,\, .\label{iz}
\end{align}
The function $I_2(z)$ comes from emission times $t\in(-\infty,\,t_s]$ corresponding to $z\in(-1,\,z_{\rm max}]$, 
\begin{align}
I_2(z)=\,&2\,I_1(z_{\rm max})-I_1(z)\label{zreversal}
\\
=&\,-R +{\textstyle\frac{1}{2}} \sqrt{\pi }\,\exp R^2\,\erf R\nonumber\\
&-(z+1)\, \sqrt{R^2-\ln ( z+1)}+{\textstyle\frac{1}{2}} \sqrt{\pi }\,\exp R^2\,\erf\sqrt{R^2-\ln ( z+1)}\,\, ,
\end{align}
where equation (\ref{zreversal}) is another manifestation of the time reversal symmetry via equation (\ref{reversal}).

An example of a Hubble diagram with red- and blueshift, but without a cusp can be found in reference \cite{st05}.

\section{Sneezing universe versus supernovae}\label{versus}

To analyze the effect of a quadratic equation of state on the evolution of the universe,
we use two data sets of type 1a supernovae. The first data set is from 
Union 2 sample \cite{union2} with 557 supernovae up to a redshift of 1.4 and the second
is from the Joint Light curve Analysis \cite{jla} with 740 supernovae up to a redshift
of 1.3. The JLA analysis simultaneously fits cosmological parameters with 
light-curve time-stretching $\alpha_s$ and colour at maximum brightness $\beta_c$.  

The following analyses use these 2 sets of supernovae independently and combined 
after statistical compatibility check. The combined sample contains 1007 
independent supernovae, but since the light curve calibrations are different
between the Union 2 and JLA analyses, two different normalization parameters $m_s$ are used. 
We use frequentist's statistics \cite{pdg} based on $\chi^2$ minimization. 
The MINUIT package  \cite{minuit} is used to find the minimum of the $\chi^2$ and 
to compute errors by using the second $\chi^2$ derivative. All our 
results are given after marginalization over the nuisance parameters ($m_s$, $\alpha_s$ and $\beta_c$).

The general $\chi^2$ is expressed in terms of the full covariance matrix of
supernovae magnitude stretch and colour including correlations and systematics provided by each
collaboration. It reads
\bb \chi^2 = \Delta M^T V^{-1} \Delta M, \ee
where  $\Delta M$  is the vector of differences between the expected supernovae 
magnitude $m_e$ and the reconstructed experimental magnitude at maximum of
the light curve $m_r$. For the JLA sample the reconstructed magnitude reads:
\bb m_r = m_{\rm peak} + \alpha_s X1 - \beta_c C, \ee
where $X1$ is related to the measured light curve time stretching, $C$ the supernovae colour 
at maximum of brightness and $m_{peak}$ the magnitude at maximum of the light
curve fit.  

The expected magnitude is written as  $m_e(z) = m_s - 2.5 \log_{10} \ell(z)$ where
$\ell(z)$ is given by the first branch of equation (\ref{luminosity}). 
Notice that the normalization parameters $m_s$ contain the unknown intrinsic
luminosity of type 1a supernovae as well as the $\tau$ parameter. 
The expected magnitude is then only a function of $m_s$, $R$ and of the redshift (\ref{iz}). 

Table~\ref{table1} presents the results of the fit using Union 2, JLA and combined
samples for a sneezing universe and the standard flat $\Lambda$CDM universe for comparison. 
The quality of the fits are identical for both models even though it is 
marginally better for the sneezing universe. This clearly indicates that the 
exotic fluid with the quadratic equation of state (\ref{state}) can replace both dark matter and dark energy at least for
supernovae data.

\begin{table}[htbp]
\begin{center}
\begin{tabular}{||c||c|c|c||c|c||} \hline
            &   \multicolumn{3}{c||}  {Sneezing Universe}  & \multicolumn{2}{c||}{$\Lambda$CDM flat Universe}  \\ \hline
            &         $R$           &$z_{\rm max}$ &  $\chi^2$            &  $\Omega_m $   &          $\chi^2$    \\ \hline
 Union 2    &  $0.99 \pm 0.03$ &$1.76\pm 0.1$    &  $530.78$         &  $0.27 \pm 0.04$  & $530.73$  \\ \hline
 JLA        &  $0.99 \pm 0.03$& $1.8\pm 0.2$    &  $701.07$         &  $0.29 \pm 0.03$  & $701.29$  \\ \hline
Combined     &  $1.00 \pm 0.03$ & $1.7\pm0.1$    &  $1003.16$        &  $0.30 \pm 0.02$  & $1003.18$  \\ \hline
\end{tabular}
\caption[]{Fit results (1$\sigma$ errors) for Union 2, JLA and combined
  samples using flat $\Lambda$CDM and sneezing universes.}
\label{table1}
\end{center}
\end{table}

\begin{figure}[h]
\begin{center}
\epsfig{figure=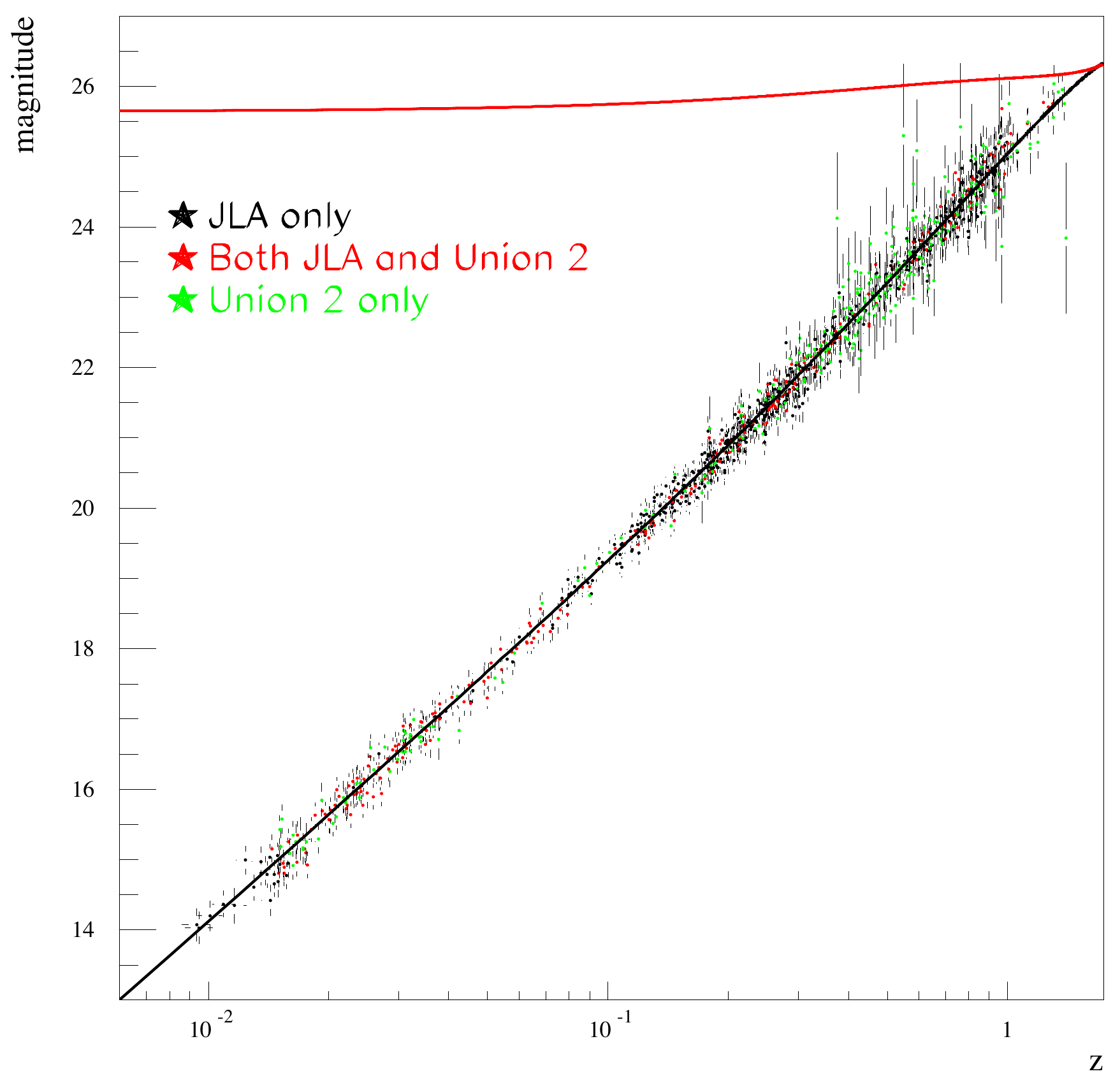,width=15cm}
\caption[]{The Hubble diagram for the three samples, Union 2, JLA and combined. The black line is the first, the red line is the second branch of the Hubble diagram computed from the sneezing universe. Only the black line was used in the fit.}
\label{figure1}
\end{center}
\end{figure}

Figure \ref{figure1} shows the data points and the black line is the best fit using only the first branch of the Hubble diagram  (\ref{luminosity}). The red line indicates the second branch. We see that the red branch is not really populated justifying somewhat {\it a posteriori}  that we have discarded it {\it a priori}. 

The  parameter $R$ for the combined sample is equal to 1 with an error of $3 \,\%$ 
and using formulae (\ref{rdef}) and (\ref{zmax}) we have a maximum redshift  of about 1.7, $z_{\rm max}=e-1$. Consequently the maximum temperature of the sneezing universe is 
\bb T_s=(z_{\rm max}+1)\, T_{\rm CMB}=
2.72\cdot2.72\,{\rm K}=7.4\,{\rm K},\ee
 at the sneeze, some 9.1 Gigayears ago.

\vfil\eject

\section{Adding baryons to the exotic fluid}

This fit encourages us to interpret the exotic fluid with its quadratic equation of state (\ref{state}) as ersatz of dark energy and dark matter together. We are therefore tempted to add ordinary matter, cold baryons with mass density $\rho _b$ and zero pressure, $p_b=0$. As we now have only one continuity equation,
\begin{align}
(\rho+\rho _b) '=-3H\,(\rho +\rho_b +p),\label{cont2}\end{align} 
 for two components, our system of differential equations is under-determined. To remain in business we cheat as is tradition by postulating two independent continuity equations for both components. These two continuity equations then integrate like charms yielding:
 \begin{align}
 \Omega _X&\dpp=\,\frac{8\pi G\,\rho }{3\,H_0^2}\,=
 \,\frac{1}{{\textstyle\frac{9}{4}}\,H_0^2\,\tau^2\,\ln(a/a_0)+1/\Omega _{X0}}\, , \label{omegax_evol}
\\[2mm]
 \Omega _b&\dpp=\,\frac{8\pi G\,\rho_b }{3\,H_0^2}\,=
\Omega _{b0} \lp\frac{a_0}{a}\rp^3. \label{omegab_evol}
\end{align} 
  From the first Friedman equation (\ref{first}) we have for the initial conditions
  \begin{align}
  \Omega _{X0}+\Omega _{b0} =1. \label{initial}
  \end{align}
  Now remains the integration of the first Friedman equation (\ref{first}) with two components. The exotic fluid alone generates the mild sneezing singularity, while the baryons alone generate the popular big bang singularity. 
  
   \begin{figure}[h]
\begin{center}
\includegraphics[width=14.5cm, height=6.5cm]{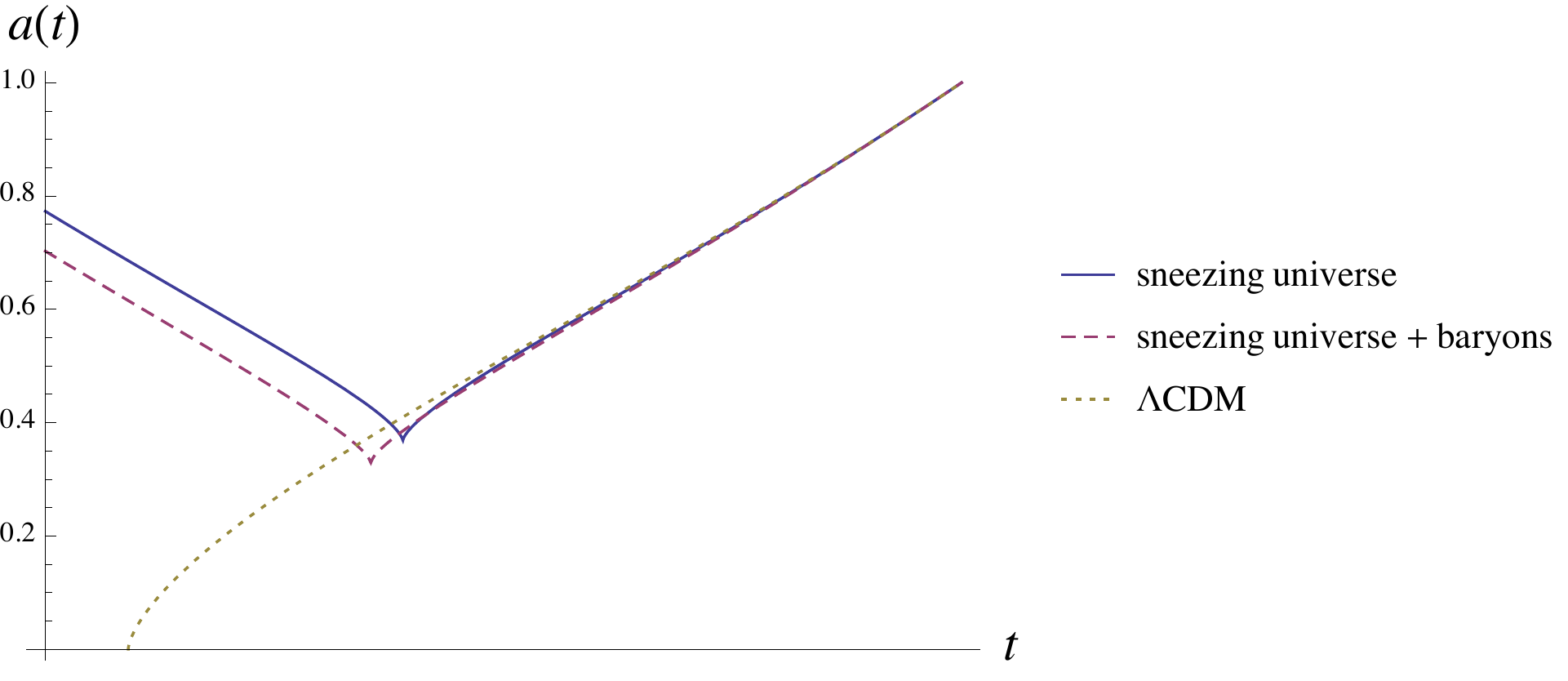}
\caption{Three scale factors\label{3a}
 }
\end{center}
\end{figure}
\begin{figure}[h]
\begin{center}
\includegraphics[width=7.5cm, height=6.5cm]{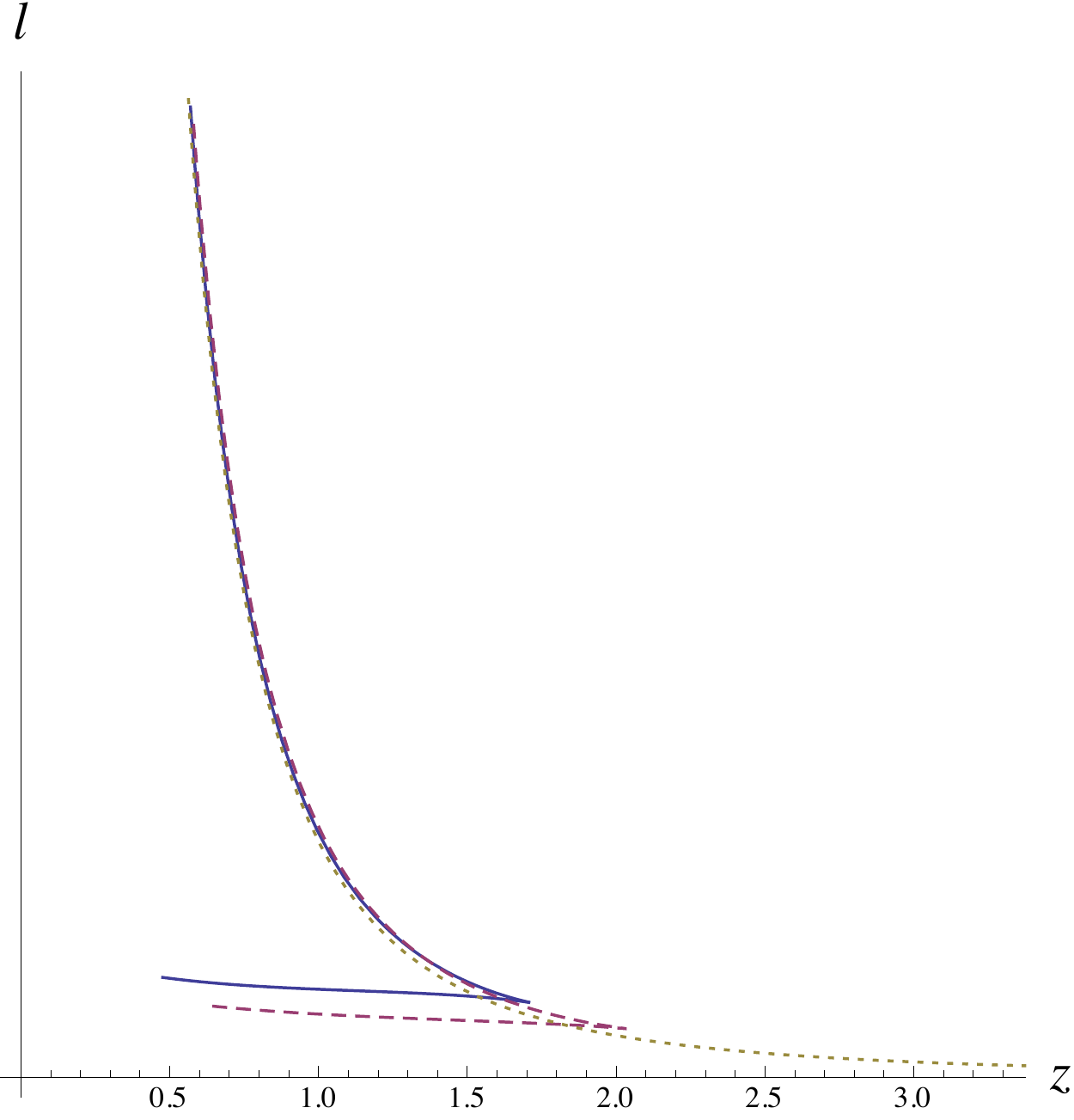}
\caption{Three Hubble diagrams, redshifts only, colour codes  as in figure \ref{3a}. \label{3ell}
 }
\end{center}
\end{figure}
  
  With both components, the first Friedman equation has still separate variables and tells us that the cold sneeze kills the hot big bang:
  \begin{align}
  \,\frac{\de a}{a\,\sqrt{[(9/4)\,H_0^2\,\tau^2\,\ln(a/a_0)+1/(1-\Omega _{X0} )]^{-1}\,+\,\Omega _{b0} ({a_0}/{a})^3}}\, =\,H_0\,\de t.\label{sep}
  \end{align}
  Indeed going back into the past starting from $t_0$, the scale factor decreases until -- at $t_s$ -- the bracket
$  [(9/4)\,H_0^2\,\tau^2\,\ln(a/a_0)+1/(1-\Omega _{X0} )]$ vanishes. At this time, the scale factor has the value
 \begin{align}
a_s\dpp=a(t_s) =a_0\,\exp-\frac{4}{9\,(1-\Omega _{b0})\,H_0^2\tau^2}\, ,
\end{align}
 and the universe sneezes and bounces. This is how it avoids the big bang. The maximum redshift,
 \begin{align}
z_{\rm max}=\,\frac{a_0}{a_s}\,-1 =\exp\frac{4}{9\,(1-\Omega _{b0})\,H_0^2\tau^2}\,-\,1,
\end{align}
  increases with respect to its value for vanishing baryon density.
We performed the integration on the left-hand side of equation (\ref{sep}) numerically.
   
   Figure \ref{3a} shows the scale factors for the $\Lambda $CDM universe, the sneezing universe and the sneezing universe with baryons added and figure \ref{3ell} shows the three corresponding Hubble diagrams, redshifts only. Again the chosen parameters anticipate results of the fit to the Hubble diagram.   
   
   Note that by the same mechanism, the sneeze continues to kill the big bang when we add radiation and spatial curvature.

      \section{Sneezing universe with a pinch of baryons versus supernovae}

We use the same kind of analysis as in section 4 to study the sneezing 
universe now with a pinch of baryons added. The only difference is in the computing
of the luminosity $\ell(z)$ from equation (\ref{luminosity}). As we have seen in the previous section,
the evolution of exotic and baryon-fluid densities in terms of the scale factor are
obtained from continuity equations and  given by formulae (\ref{omegax_evol}) and (\ref{omegab_evol}). 
The first Friedman equation (\ref{first}) is solved with these two independent 
components using the Rugge-Kutta algorithm \cite{rk} with a step in time 
corresponding to an equivalent step in redshift well below the 
experimental redshift error ($10^{-3}$) and by assuming again that we only see supernovae that exploded after the sneeze. The initial conditions are $a(0) = a_0 = 1$ s, $\Omega_X(0) = \Omega_{X0}$ and 
$\Omega_b(0) = \Omega_{b0}$ with the usual closure relation (\ref{initial}) for a flat universe. 

The final fit procedure is then only a function of $m_s$, 
time-stretching correction $\alpha_s$, colour correction $\beta_c$,
$\Omega_{b0}$ and the new dimensionless parameter $\tau H_0$ describing 
the weight of the quadratic term in the equation of state (\ref{state}). 
Notice the simple relation between $\tau H_0$ and $R$: $\tau H_0 = 2/3  \cdot R^2$ when $\Omega _{b0}=0$.

\begin{table}[htbp]
\begin{center}
\begin{tabular}{||c|c|c|c|c|c|c||} \hline
            & $\Omega_{sneeze}$ & $\Omega_{baryons}$ &  $\tau H_0$   &    $R$    &   $z_{\rm max}$   &   $\chi^2$    \\ \hline
 Union 2    &$ 1.0^{+0.}_{-0.15}$   &$0.^{+0.15}_{-0.}$    &$0.66^{+0.02}_{-0.66}$& $0.99$&  $1.77$ &  $530.78$   \\ \hline
 JLA        &$ 0.85^{+0.14}_{-0.15}$&$0.14^{+0.15}_{-0.14}$&$0.60^{+0.06}_{-0.6}$&$0.95$ &  $3.2 $ &  $700.33$   \\ \hline
Combine     &$ 0.97^{+0.02}_{-0.3}$&$0.02^{+0.3}_{-0.02}$&$0.67^{+0.01}_{-0.67}$&$1.00$&  $1.74$ &  $1003.1$   \\ \hline
\end{tabular}
\caption[]{Fit results (1$\sigma$ errors) for Union 2, JLA, and combined samples using the sneezing universe with some cold matter.}
\label{table2}
\end{center}
\end{table}

Table \ref{table2} shows our results of the fit for the three samples 
of supernovae. Despite the fact that we have one more free parameter, 
the quality of the fit is only marginally improved compared to a pure 
sneezing universe. However, the baryon density we find is statistically
compatible with the generally admitted value of about $4\%$. The  parameter $R$ is 
very close to 1 as before and the maximum redshift is still quite small
(of the order of 1.7) excepted for the JLA sample which gives $z_{\rm max}= 3.2$
because of the bigger amount of baryon density ($14\%$).

\begin{figure}[h]
\begin{center}
\epsfig{figure=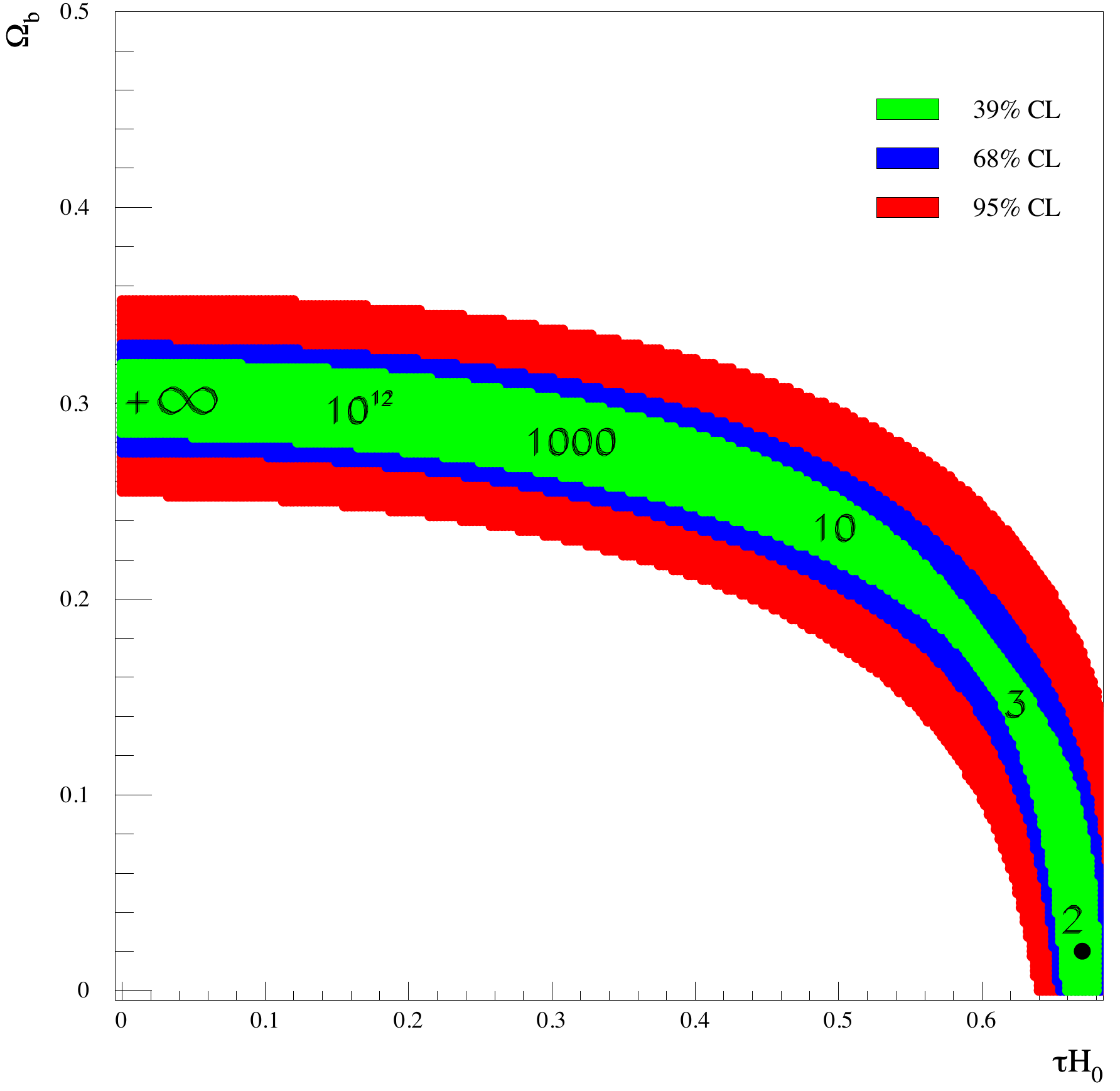,width=15cm}
\caption[]{Confidence levels for the sneezing universe with cold matter. The black dot
  represents the minimum $\chi^2$. The numbers indicate the maximum redshifts $z_{\rm max}$.}
\label{figure2}
\end{center}
\end{figure}

Figure \ref{figure2} shows the confidence level contour on 
$\Omega_{b0}$ versus $\tau H_0$. As expected, the degeneracy between 
both fluids is important and explains the big errors in Table \ref{table2}. The best fit is indicated by the black dot, The numbers give the maximum redshifts. At a confidence level better than $36\%$,  
both dark matter and dark energy can be replaced by the exotic fluid with a pinch of baryons.   
\vfil\eject

\section{Conclusions}

Just as dark energy with $p=w\,\rho $, the quadratic equation of state (\ref{state}) is a 1-parameter extension of the cosmological constant. However the two ensuing universes differ dramatically: the second one has no big bang, except if $\tau=0$.

To fit the supernova data, dark energy needs the addition of cold dark matter and then retrieves the standard model. A nice feature of the sneezing universe is that it does not need cold dark matter to produce an excellent fit. Any amount of baryons, $\Omega _{b0}$, may even be added without spoiling this fit. A further remarkable property of the fit in absence of baryons is that with a precision of 3 \% the characteristic time scale $\tau$ in the quadratic equation of state (\ref{state}) coincides with the time span $t_0-t_s$ between the mild singularity, the `sneeze', and today. 
We finally note that, thanks to its mild bounce without big crunch, the sneezing universe is an example of a ``New Ekpyrotic Cosmology'' \cite{buchbinder}, a possible alternative to inflationary models.

However the sneezing universe has also dangerous properties: 
\begin{itemize}\item
It has no room for objects with redshift exceeding 
\bb z_{\rm max}= \exp\frac{4}{9\,(1-\Omega _{b0})\,H_0^2\tau^2}\,-\,1,\nonumber\ee
that is
 $z_{\rm max}=1.7$ without baryons.
\item
The temperature of the universe is always lower than its temperature at sneeze,\\ $(z_{\rm max}+1)\cdot T_{\rm CMB}$, that is 7.4 K without baryons. Therefore generically, it never ionized and was  always transparent.
\item
Without baryons the sneeze happened only some 9.1 Gigayears ago.
\item
The sneezing universe predicts a cusp in the Hubble diagram at rather low redshift and a second branch coming from supernovae that exploded before the sneeze. This second branch is absent from the data.
\item
This second branch also predicts supernovae with blueshift and high apparent luminosity.
\item
Olbers' paradox seems to make a come back in the sneezing universe.
\end{itemize}

Certainly one can try and invoke new physics around the sneeze in order to avoid these dangerous properties, but we think that the sneezing universe deserves a cleaner dismissal. 

In order to confront the sneezing universe with Large Scale Structure data or lensing data, we need to know how the quadratic equation of state (\ref{state}) modifies the Schwarzschild solution. This question is currently under investigation.

However we think that the cleanest dismissal comes from a confrontation with Cosmic Micro-wave Background data and we urge our colleagues practicing cosmic perturbation theory to take up this challenge. \\
${}$\vspace{6mm}\\
\noindent
{\bf Acknowledgements:} This work has been carried out thanks to the support of the OCEVU Labex
(ANR-11-LABX-0060) and the A*MIDEX project (ANR-11-IDEX-0001-02) funded
by the "Investissements d'Avenir" French government program managed by
the ANR.

\end{document}